\documentclass[varenna]{cimento}

\usepackage{graphicx}  

\title{New Results from MiniBooNE: A Search for Electron Antineutrino Appearance at $\sim$1 eV$^2$}

\author{G.~Karagiorgi, for the MiniBooNE Collaboration}

\institute{Massachusetts Institute of Technology, Cambridge, MA 02139, USA}

\PACSes{14.60.Lm, 14.60.Pq, 14.60.St}

\begin{document}

\maketitle

\begin{abstract}
These proceedings summarize the first MiniBooNE electron antineutrino appearance search results, corresponding to a data sample collected for 3.39$\times$10$^{20}$ protons on target (POT). The search serves as a direct test of the LSND oscillation signature, and provides complementary information which can be used in studies addressing the MiniBooNE neutrino-mode low-energy excess.
\end{abstract}

\section{The MiniBooNE and LSND anomalies}
The MiniBooNE experiment has performed a search for $\bar{\nu}_{\mu}\to\bar{\nu}_e$ oscillations at large $\Delta m^2$ \cite{mbnubar}, an oscillation signature that had been observed by the LSND experiment, with 3.8$\sigma$ significance \cite{lsnd}. This oscillation interpretation relies on the existence of a fourth, sterile neutrino mass eigenstate, with $\Delta m^2\sim$ 0.1-100 eV$^2$. Mixing via this fourth mass eigenstate can lead to a small probability amplitude for $\bar{\nu}_{\mu}\to\bar{\nu}_e$ and $\nu_{\mu}\to\nu_e$ oscillations at $L[m]/E[MeV]\sim$ 1. MiniBooNE has previously searched for this type of oscillation using a neutrino beam \cite{mbnu}, and, under the assumption of CPT conservation, has excluded the LSND interpretation 98\% confidence level (CL) \cite{mbnu}. At the same time, the search revealed an excess of $\nu_e$ events at low energy \cite{mblowe}, which is inconsistent with the LSND excess under the a single sterile neutrino oscillation hypothesis; however, extensions of this model \cite{sterile} offer the possibility of reconsiling the MiniBooNE neutrino results with the LSND antineutrino result. These models involve large CP violation which leads to different probabilities for $\nu_{\mu}\to\nu_e$ as opposed to $\bar{\nu}_{\mu}\to\bar{\nu}_e$ oscillations. Other models \cite{explanations} have also been suggested as explanations, some of which offer predictions for antineutrino running at MiniBooNE. In order to provide another handle on the low-energy excess, MiniBooNE was approved in 2007 for extended antineutrino running \cite{mbloi}, which also enabled MiniBooNE to perform a direct test of the LSND oscillation interpretation, using antineutrinos.

\section{The MiniBooNE Electron Antineutrino Appearance Search}
The MiniBooNE experiment uses 8 GeV protons incident on a beryllium target in order to produce mesons which subsequently decay to generate the neutrino beam. A magnetic field is used at the target to focus positively charged mesons in the forward direction, and defocus negatively charged mesons, to produce a neutrino beam. Reversing the polarity of the magnetic field allows MiniBooNE to switch from a neutrino to an antineutrino beam. The flux \cite{mbflux} consists primarily of $\bar{\nu}_{\mu}$ and $\nu_{\mu}$. The low $\nu_e$ and $\bar{\nu}_e$ content of the beam minimizes the background to the oscillation search, ensuring sensitivity to small-amplitude (of order 10$^{-3}$) oscillations. The $\bar{\nu}_{\mu}$ flux has a mean energy of $\sim$600 MeV. The MiniBooNE detector \cite{mbdetector} is located at L=541 m from the proton target. This establishes an $L/E$ similar to LSND, and therefore sensitivity to $\Delta m^2\sim$ 1 eV$^2$. The detector is a spherical tank, 12 meters in diameter, filled with mineral oil and lined with photomultiplier tubes (PMTs). The particle detection and identification method relies on the detection of cherenkov and scintillation light emitted by outgoing charged particles which are produced in neutrino interactions. 

The antineutrino oscillation analysis \cite{mbnubar} employs the same analysis chain that was implemented in neutrino mode \cite{mblowe}. The analysis relies on differentiation between a majority of $\bar{\nu}_{\mu}$ charged-current quasi-elastic (CCQE) events, and $\bar{\nu}_e$ CCQE events, which are the signal. A track-based event reconstruction is implemented, which uses PMT hit topology and timing to identify electron-like or muon-like cherenkov rings from the corresponding CCQE interactions. The $\bar{\nu}_e$ and $\bar{\nu}_{\mu}$ CCQE spectra are fitted simultaneously as a function of reconstructed antineutrino energy, $E^{QE}_{\nu}$, and the oscillation parameters $\Delta m^2$ and $\sin^2 2\theta$ are extracted. The $\bar{\nu}_{\mu}$ CCQE prediction is used in the fit in order to provide a constraint to the $\bar{\nu}_{e}$ CCQE prediction, as both spectra are correlated through flux and cross section systematics.

\begin{figure}
\begin{center}
\vspace{1cm}
\includegraphics[angle=-90, trim=100 0 0 0, width=9cm]{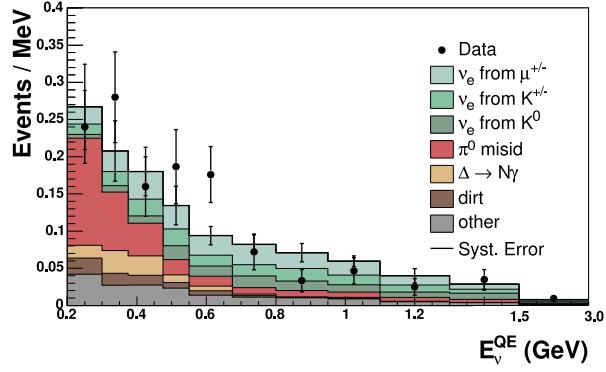}     
\end{center}
\caption{\label{fig:nue} MonteCarlo background prediction and observed data as a function of reconstructed neutrino energy for the present antineutrino data sample corresponding to 3.39$\times$10$^{20}$ POT.}
\end{figure}

\begin{figure}
\begin{center}
\includegraphics[trim=0 50 0 0,width=7cm]{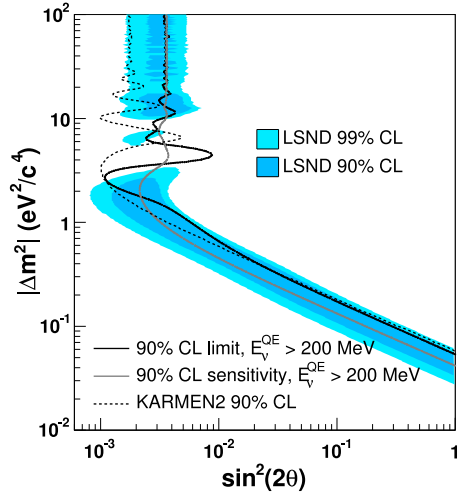}     
\end{center}
\caption{\label{fig:lim} The antineutrino sensitivity and limit to LSND-allowed $\bar{\nu}_{\mu}\to\bar{\nu}_e$ oscillations, from fits to 200 $<E^{QE}_{\nu}<$ 3000 MeV. The MiniBooNE antineutrino dataset corresponds to 3.39$\times$10$^{20}$ POT. Also shown is the limit from KARMEN \cite{karmen}.}
\end{figure}

The $\bar{\nu}_e$ CCQE background prediction, for 3.39$\times$10$^{20}$ POT, is shown in fig.~\ref{fig:nue}. The background is dominated at low energy by mis-identified $\bar{\nu}_{\mu}$ events, such as neutral-current (NC) $\pi^0$ interactions. That is because MiniBooNE cannot differentiate between a single photon or a single electron produced in the detector. At high energy, the dominant background is CCQE interactions of intrinsic $\bar{\nu}_e$ from the beam, produced in $K$ and $\pi\to\mu$ decays. The sensitivity to $\bar{\nu}_{\mu}\to\bar{\nu}_e$ oscillations is shown in fig.~\ref{fig:lim}. The MiniBooNE sensitivity provides substantial coverage of the lower $\Delta m^2$ region allowed at 90\% CL by a joint analysis of LSND and KARMEN \cite{karmen} data \cite{mbnubar}.

\section{Results}
The reconstructed energy spectrum of $\bar{\nu}_e$ CCQE data is shown in fig.~\ref{fig:nue}, overlaid on the predicted $\bar{\nu}_e$ CCQE background. At energies above 475 MeV, the data agree with the background prediction within systematic and statistical uncertainties. The 475-675 MeV energy region shows a 2.8$\sigma$ data fluctuation above background prediction. This fluctuation forces the MiniBooNE limit, shown in fig.~\ref{fig:lim}, to be significantly worse than the sensitivity at lower $\Delta m^2$. The MiniBooNE best oscillation fit corresponds to ($\Delta m^2=$ 4.4 eV$^2$, $\sin^2 2\theta=$ 0.004).

Interestingly, the low-energy region (200-475 MeV) shows no evidence of an excess. In this range, MiniBooNE observes 61 events, compared to a background expectation of 61.5$\pm$11.7 (sys+stat) events. Table \ref{tab:excess} shows the probability (from a two-parameter fit to the data) that each of the following hypotheses explains the observed number of low-energy neutrino and antineutrino events: 
1) Same $\sigma$: Same NC cross section for neutrinos and antineutrinos. 2) $\pi^0$ scaled: Scaled to the number of NC $\pi^0$ events. 3) POT scaled: Scaled to number of POT. 4) BKGD scaled: Scaled to the total number of background events. 5) CC scaled: Scaled to the number of CC events. 5) Kaon scaled: Scaled to the number of low-energy $K$ events. 6) Neutrino scaled: Scaled to the number of neutrino events. The same $\sigma$, POT scaled, and Kaon scaled hypotheses are disfavored as explanations of the MiniBooNE low-energy excess. The most preferred model is that where the low-energy excess is contributed from only neutrinos in the beam.

\begin{table}
  \caption{The $\chi^2$-probability that each hypothesis explains the observed number of low-energy neutrino and antineutrino events, assuming only statistical, fully correlated systematic, and fully uncorrelated systematic errors. A proper treatment of systematic correlations is in progress.}
  \label{tab:excess}
  \begin{tabular}{lccc}
    \hline
Hypothesis & stat.-only & stat. and correlated sys. & stat and uncorrelated sys. \\
\hline
Same $\sigma$ & 0.1\% & 0.1\% & 6.7\% \\
$\pi^0$ scaled & 3.6\% & 6.4\% & 21.5\% \\
POT scaled & 0.0\% & 0.0\% & 1.8\% \\
BKGD scaled & 2.7\% & 4.7\% & 19.2\% \\
CC scaled & 2.9\% & 5.2\% & 19.9\% \\
Kaon scaled & 0.1\% & 0.1\% & 5.9\% \\
$\nu$ scaled & 38.4\% & 51.4\% & 58.0\% \\
\hline
\end{tabular}
\end{table}

\section{Conclusion}
MiniBooNE has performed a blind analysis for $\bar{\nu}_{\mu}\to\bar{\nu}_e$ oscillations. The $\bar{\nu}_e$ data is found in agreement with the background prediction as a function of $E^{QE}_{\nu}$. No strong evidence for oscillations in antineutrino mode has been found, given the current statistics. Interestingly, there is no evidence of significant excess at low energy in antineutrino mode. This has already placed constraints to various suggested low-energy excess interpretations. MiniBooNE is currently collecting more antineutrino data, for a total of 5.0$\times$10$^{20}$ POT, and has been approved for further running, to collect a total of 10.$\times$10$^{20}$ POT. This will improve sensitivity to oscillations, and allow further investigation of the neutrino-mode low-energy excess. Additional information will be provided by the NuMI-beam neutrinos detected at MiniBooNE \cite{numi}.


\begin{thebibliography}{0}

\bibitem{mbnubar}
  \BY{Aguilar-Arevalo~A.~A. {\em et al.}}
  preprint arXiv:0904.1958 [hep-ex].
\bibitem{lsnd}
  \BY{Athanassopoulos~C. {\em et~al.}}
  \IN{Phys. Rev. Lett.}{75}{1995}{2650};
  \SAME{77}{1996}{3082};
  \SAME{81}{1998}{1774};
  \BY{Aguilar-Arevalo~A.~A. {\em et~al.}}
  \IN{Phys. Rev. D}{64}{2001}{112007}.
\bibitem{mbnu}
  \BY{Aguilar-Arevalo~A.~A. {\em et~al.}}
  \IN{Phys. Rev. Lett.}{98}{2007}{231801}.
\bibitem{mblowe}
  \BY{Aguilar-Arevalo~A.~A. {\em et~al.}}
  \IN{Phys. Rev. Lett.}{102}{2009}{101802}.
\bibitem{sterile}
  \BY{Sorel~M. {\em et~al.}}
  \IN{Phys. Rev. D}{70}{2004}{073004};
  \BY{Maltoni~M. \atque Schwetz~T.}
  \IN{Phys. Rev. D}{76}{2007}{093005};
  \BY{Karagiorgi~G.}
  \IN{AIP Conf. Proc.}{981}{2008}{210}.
\bibitem{explanations}
  \BY{Harvey~J.~A. {\em et~al.}}
  \IN{Phys. Rev. Lett.}{99}{2007}{261601};
  \SAME{Phys. Rev. D}{77}{2008}{085017};
  \BY{Pas~H. {\em et~al.}}
  \IN{Phys. Rev. D}{72}{2005}{095017};
  \BY{Goldman~T. {\em et~al.}}
  \IN{Phys. Rev. D}{75}{2007}{091301};
  \BY{Nelson~A.~E. \atque Walsh~J.}
  \IN{Phys. Rev. D}{77}{2008}{033001};
  \BY{Kostelecky~V.~A. \atque Mewes~M.}
  \IN{Phys. Rev. D}{69}{2004}{016005};
  \BY{Katori~T. {\em et~al.}}
  \IN{Phys. Rev. D}{74}{2006}{105009};
  \BY{Giunti~C. \atque Laveder~M.}
  \IN{Phys. Rev. D}{77}{2008}{093002}.
\bibitem{mbloi}
  \BY{Aguilar-Arevalo~A.~A. {\em et~al.}}
  \TITLE{Addendum to the MiniBooNE Run Plan: MiniBooNE Physics in 2006} 2006.
\bibitem{mbflux}
  \BY{Aguilar-Arevalo~A.~A. {\em et al.}}
  \IN{Phys. Rev. D}{79}{2009}{072002}.
\bibitem{mbdetector}
  \BY{Aguilar-Arevalo~A.~A. {\em et~al.}}
  \IN{Nucl. Instrum. Meth.  A}{599}{2009}{28}.
\bibitem{karmen}
  \BY{Armbruster~B. {\em et~al.}}
  \IN{Phys. Rev. D}{65}{2002}{112001}.
\bibitem{numi}
  \BY{Adamson~P. {\it et al.}}
  \IN{Phys. Rev. Lett.}{102}{2008}{211801}.

\end{thebibliography}
\end{document}